\documentclass{ismdproc}
\def\epem{$e^+e^-$}
\def\ednd3p{E\frac{dN}{d^3p}}
\def\edsd3p{E\frac{d\sigma}{d^3p}}

\def\lapproxeq{\lower .7ex\hbox{$\;\stackrel{\textstyle
<}{\sim}\;$}}
\def\gapproxeq{\lower .7ex\hbox{$\;\stackrel{\textstyle
>}{\sim}\;$}}
\begin{document}
\title{Universal properties of particle production\\
 in the soft limit $p_T\to 0$}

\author{{\slshape Wolfgang Ochs$^1$\footnote{Speaker}, Valery A. Khoze$^2$
and    M.G. Ryskin$^3$}\\[1ex]
$^1$Max Planck Institut f\"ur Physik, Werner-Heisenberg-Institut,
D-80805 Munich, Germany\\
$^2$Institute for Particle Physics Phenomenology, University of Durham,
DH1 3LE, UK \\
$^3$Petersburg Nuclear Physics Institute, Gatchina, St.~Petersburg,           
188300, Russia}

%

\contribID{xy}  
\confID{yz}
\acronym{ISMD2010}
\doi            

\maketitle

\begin{abstract}
The momentum spectra of particles in different high energy processes,
such as $e^+e^-$ annihilation, $pp$ and nuclear collisions 
in the limit $p,p_T\to 0$ exhibit similar properties because of the dominant
role 
of coherent soft gluon bremsstrahlung. 
We observe the following general features:
the inclusive particle density
approaches a limiting behaviour and becomes independent of primary collision 
energy;
furthermore, 
it becomes proportional to the QCD colour factors $C_A,C_F$ which appear in
the Born term for the respective minimal partonic processes. 
In this limit, nuclear collisions reach with good accuracy 
participant (``wounded nucleon'') scaling.
Particle ratios in the low momentum region display
a universal behaviour. Future measurements at
the LHC will provide crucial tests for 
the contributions from additional incoherent
multi-component processes.
\end{abstract}

\section{Limiting soft particle emission}
We consider the invariant hadron density $\ednd3p$ in the
limit $p_T\to 0$ or $p\to 0$ and we define 
\begin{equation}
 I_0=\left. E\frac{dN}{d^3p}\right|_{p\to 0}. \label{i0}
\end{equation}
If we  calculate the momentum spectrum of gluons within perturbative 
QCD we find that the lowest order contribution,
i.e. the Born term, dominates in the soft limit. 
This leads to some universal features
for the various processes although they show very different 
features with increasing complexity in the multi-particle final states:
quark and gluon jet production in \epem,
spectator jets and underlying event in $pp$ and collective phenomena 
in $AA$ collisions. Nevertheless, in the soft
limit, we expect for the final state gluons the following properties:\\
1.~momentum spectra become energy independent;\\
2.~the relative normalisation of spectra in different processes is given by
the colour factors which appear in the Born term for the respective 
minimal partonic process. \\
The arguments hold for partons and we assume 
these properties are also valid for hadrons.

First, we present qualitative arguments for this behaviour. A soft gluon is
coherently emitted by all final state partons. Due to its large
wavelength it cannot resolve the intrinsic structure of the bunches of
partons and so it ``sees'' only the overall colour charge of the
primary partons.

The details of this study with the emphasis on $pp$ and $AA$ results 
are presented in Ref. \cite{Ochs:2010sv}.
The corresponding properties for \epem annihilation were obtained already 
some time ago \cite{Khoze:1996ij} (for a recent overview, see
\cite{Ochs:2010vy}).

The inclusive spectra can be derived using evolution equations, similar to
the well known DGLAP equation, but with modification at small momenta to
include coherence effects of soft gluons (for a review, see e.g.
Ref. \cite{Khoze:1996dn}). In the simplest double logarithmic 
approximation with fixed coupling $\alpha_s$ 
one finds for the inclusive distribution in angle $\Theta$ and energy $k$ 
the perturbative expansion
\begin{equation}
 \displaystyle     
\frac{dN_a}{dk d\Theta} = \frac{2}{\pi}\frac{C_a}{k \Theta}\alpha_s
 + \frac{4N_C}{\pi^2}\frac{C_a}{k \Theta}\alpha_s^2  
    \ln\frac{E}{k}\ln \frac{k_T}{Q_0}
+\ldots \label{Dborn}
\end{equation}
The first term on the r.h.s. represents 
the well known Born expression for the
soft bremsstrahlung which is independent of jet energy $E$ and holds the 
colour factor 
for the quark or gluon jet, respectively, with  $C_a=C_F$ and $C_a=C_A$.
This term dominates if the transverse momentum $k_T\approx
k\Theta$ approaches the lowest value at the cut-off $Q_0$, and this property
remains valid in more exact calculations. The application to hadrons assumes
similarity of parton and hadron spectra as required by
the LPHD concept~\cite{Azimov:1984np}
with small $Q_0$ of few 100 MeV and accounting for the 
mass effects. 
Then
a good description of the \epem data in a wide range of energies
$\sqrt{s}=3\ldots 200$ GeV is obtained demonstrating the approach to a
limiting behaviour for $p\to 0$.

The difference between 
quark and gluon jets and the appearance of colour factors
in the Born term (\ref{Dborn}) is best demonstrated in the measurement of
soft radiation in 3-jet events perpendicular to the jet 
plane for which case the formulae for particle densities 
are given in~\cite{Khoze:1996ij}. 
Varying the
inter-jet angles one can interpolate between the collinear configurations
corresponding to $q\bar q$ and  $gg$ dipoles. This allows a test of the
prediction
\begin{equation}
I_0^{gg}/I_0^{q\bar q}= C_A/C_F. \label{rgq}
\end{equation}
The measurement by DELPHI
\cite{Abdallah:2004uu} confirmed this expectation
and allowed a determination of the ratio in (\ref{rgq}) as $2.211\pm 0.053$
consistent with $C_A/C_F=9/4$.
In contrast, the ratio of total multiplicities in quark and gluon 
jets acquires large corrections to the ratio
$C_A/C_F$.

\section{High-energy $pp$ collisions} 

We consider next the soft particle production in the non-diffractive
``minimum bias'' events in $pp$ collisions.  As in \epem annihilation we
look for the minimal partonic process which could be responsible for the
very soft gluon bremsstrahlung.  We assume that the minimal process
of lowest order corresponds to the semihard 
one-gluon exchange between any two partons in the proton with 
a dominantly small scattering angle and a non-vanishing cross section at
high energies. The exchange of a gluon between the partons in the proton 
leads to two separating 
outgoing partonic systems which give rise to gluon bremsstrahlung from
the effective colour octet dipole. In the simplest case each proton splits into
a quark-diquark pair which scatter via one-gluon exchange. 
Also more complex partonic processes could occur 
with two colour octet systems, as discussed 
in  Ref. \cite{Ochs:2010sv}.

Therefore, we expect  limiting behaviour
of the soft particle density in $pp$ collisions, $I_0^{pp}$, for $p\to 0$ or
$p_T\to 0$ at fixed rapidity $y$. 
Furthermore, we predict the ratio to 
the corresponding density in \epem annihilations 
\begin{equation}
p\to 0:\qquad   I_0^{pp}/I_0^{e^+e^-}\approx C_A/C_F, \label{ratio94}  
\end{equation}   
similar to the case of primary $gg$ and $q\bar q$ dipoles.
The $p_T$
distributions in both processes differ and, therefore, the integrated
multiplicities differ as well. 

In order to test these predictions we studied 
the invariant cross sections $\edsd3p$ measured in the energy range
$\sqrt{s}=20\ldots 1800$ GeV at the colliders at BNL, CERN and Fermilab.
After normalisation to the inelastic cross section, $\sigma_{in}$, we found
the corresponding density $I_0$ from extrapolation of the $p_T$ spectra to $p_T=0$.
Most experimental groups fitted their data using
parametrisations with the
small $p_T$ behaviour
\begin{equation}
\edsd3p= A \exp(B p_T +\ldots), \label{invfit}
\end{equation}
which work well down to the smallest measured values of $p_T\sim 0.1$ GeV.
We derive $I_0^{pp}=A/\sigma_{in}$ from these fits.
The functional form (\ref{invfit}), however, is not analytic at $p_T=0$.
There is no such problem from the theoretical point of view for the
``thermal'' parametrisation in terms of the transverse mass 
$m_T$ instead of $p_T$
\begin{equation}
\edsd3p= \frac{A}{(\exp(m_T/T)-1)};\quad m_T=\sqrt{m^2+p_T^2}.
\label{invfitth}
\end{equation}
This form was applied by the PHOBOS collaboration \cite{phobos}, 
and they obtained
a good fit to their $AuAu$ data down to $p_T\sim 0.03$ GeV. With such
extrapolation the resulting value of $I_0^{pp}$ is found $\sim$25\% lower
as compared to the fit with parametrisation (\ref{invfit}).

\begin{figure}[t]
\begin{center}   
\mbox{\epsfig{file=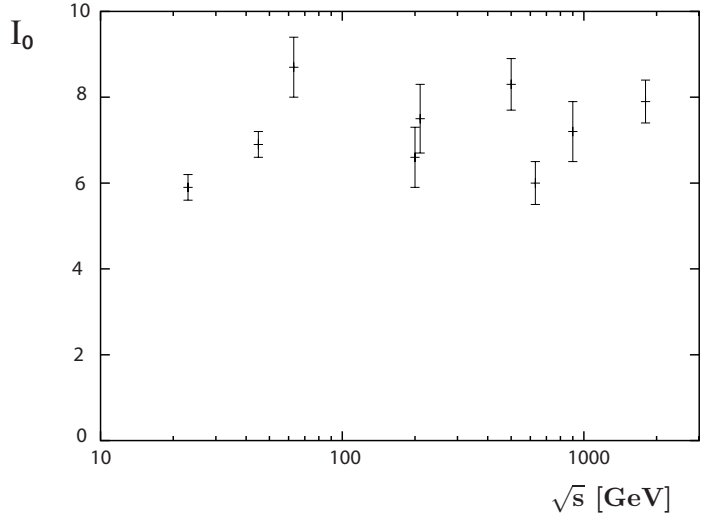,angle=0,bb=0 30 510 380,clip=,width=10cm}}
\end{center}
\vspace{-0.5cm}
\hspace*{10.0cm}{\bf ${\bf \sqrt{s}}$ [GeV] }
\caption{Soft limit $I_0^{pp}$ of the invariant density $E\frac{dn}{d^3p}$
of charged particles [$(h^+ +h^-)/2$] in $pp$ collisions as a function of 
c.m.s. energy $\sqrt{s}$ (for exponential extrapolation, normalised to
$\sigma_{in}$).}
\label{fig:I0}
\end{figure}

The results from the available published
exponential extrapolations and normalisation to inelastic cross sections 
$\sigma_{in}$ are shown in Fig. \ref{fig:I0}.
The data on $I_0^{pp}$ are seen to be consistent with energy independence over 
two orders of magnitude 
in energy $\sqrt{s}$ with fluctuations around a mean value 
$I_0^{pp}\approx (7\pm1)$ GeV$^{-2}$. 
The different functional forms of extrapolations mentioned above
do not affect this conclusion~\cite{Ochs:2010sv}. 
It should be noted that the $p_T$-integrated 
rapidity density $\frac{dN}{dy}$ doubles in this energy range. 

The observation of energy independence is similar to that in \epem
annihilation and corresponds to the expectation from
coherent bremsstrahlung. In contrast, for an incoherent superposition
of several sources one would expect a rising $I_0^{pp}$, since the number of
such processes would typically increase with energy. It will be interesting
to obtain corresponding results from the LHC at the higher energies to test this
possibility. Some results were obtained by CMS~\cite{cms} 
which show the convergence of spectra at different energies
towards $p_T=0$.

Since the soft gluon emission is driven by the lowest order
diagram 
we  predict
the same limiting behaviour 
($d\sigma/dyd^2p_T\to const$ in $\sqrt{s}$ at $p_T\to 0$)
to be observed
at any rapidity and not {\it only} at $y=0$ and we
expect a flat rapidity plateau in this limit. 
This is true, as long as we can neglect the
kinematical boundary effects for large rapidities $y$.
To study the rapidity dependence it is important to perform calculations 
with the
exact kinematics using particle masses. The use of pseudorapidity 
$\eta=-\ln\tan (\theta/2)$ will lead to a different spectral shape in $\eta$
and $p_T$ as discussed in detail in Ref. \cite{sheg}.
 
For comparison of particle densities in $pp$ and \epem collisions
in order to test eq. (\ref{ratio94}) it is better to use for normalisation 
the
non-diffractive cross section which we take as being 15\% lower than 
the inelastic
one, and we obtain 
$I^{pp}_{0,nd}\approx (8\pm 1)$~GeV$^{-2}$\cite{Ochs:2010sv}. Then we find
for the non-diffractive (minimum bias) events for a thermal fit the reduced
value 
\begin{equation}
I^{pp}_0\approx (6\pm 1)\ {\rm GeV}^{-2}.
\label{I0pp}
\end{equation}
In order to compare with \epem annihilation let us consider two results.

1. The TPC/2$\gamma$ collaboration \cite{tpc} 
directly compared their own data on
\epem annihilation with the $pp$ data by the 
BS~collaboration~\cite{Alper:1975jm}. The $p_T$ spectrum in the $pp$ collision 
is found to be steeper than
that in \epem annihilation , which is  determined with respect to the sphericity
axis. The extrapolation down to the smallest $p_T$ yields a larger
density in $pp$ collisions by factors 2.0-2.7 depending on the type of the fit.

2. Other \epem \ experiments presented cross sections as a 
function of particle energy. Using their exponential fits in this variable
in the energy range 
$\sqrt{s}=10\ldots 29$ GeV we find
$I_0^{e^+e^-}\approx (3.3\pm0.5)$ GeV$^{-2}$ or the ratio
\begin{equation}
I_0^{pp} / I_0^{e^+e^-}  \approx (1.8\pm 0.4) \div  (2.4\pm 0.5),
\text{          }\label{I0pe}
\end{equation} 
where the first (preferred) number refers to the thermal and the second to the
exponential extrapolation. This result is consistent with our 
expectation for this ratio $C_A/C_F=2.25$. The approximate energy
independence of the quantity $I_0$ in both collision processes and that their
ratio is close to 2 are remarkable and represent a serious argument in favour of
the dominance of the elementary bremsstrahlung process,
also in soft $pp$ collisions.

\section{Low $p_T$ spectra in nuclear collisions}
For nuclear collisions the data are often presented in
terms of a corresponding
ratio to the $pp$ collision according to the two limiting cases:

1. We can define the ``nuclear modification factor''
\begin{equation}
R_{AA}^{N_{coll}}=\frac{1}{N_{coll}}\frac{dN_{AA}/dp_T}{dN_{pp}/dp_T},
\label{RAB}
\end{equation}
which is equal to unity in the case of a pointlike interaction. Here the number of
nucleon-nucleon collisions $N_{coll}$ can be calculated from the Glauber model.

2. Another limiting case applies to the bulk particle production
without any hard scale. In such a case we expect that
soft particles are produced coherently over a certain range $r$ 
characteristic for
a nucleon (or even nucleus) which results in a reduced rate.
Indeed, all collaborations at RHIC
\cite{phobos,star,phenix,brahms} find that the ratio $R_{AA}^{N_{coll}}$
is falling below unity for small $p_T$. 
Data in this kinematic region are conveniently presented 
by normalising to the number of ``participating nucleons'' 
\begin{equation}
R_{AA}^{N_{part}}=\frac{1}{(N_{part}/2)}\frac{dN_{AA}/dp_T}{dN_{pp}/dp_T}.
\label{RABpart}   
\end{equation}
Bialas, Bleszynski and Czyz \cite{bialas} have introduced the concept of
participating nucleons, or ``wounded nucleons'' in their terms, and 
observed that
$R_{AA}^{N_{part}}\approx 1$ for soft production, which means that
each interacting nucleon
should be counted only once and rescatterings be disregarded. 

As before, we consider first 
the energy dependence of the particle density at
$p_T\to 0$. 
The variation of $p_T$ spectra between the two energies 62.4 and 200 GeV  
was studied for different centralities by PHOBOS~\cite{phobos}. 
The nuclear modification factors
$R_{AA}^{N_{coll}}$ approach about the same values for the two
energies at small $p_T\sim  0.2$ GeV, and the same is true for 
$R_{AA}^{N_{part}}$. This implies that the energy
dependence of $AA$ data is as weak as that of $pp$ data.

Next we investigate the normalisation of $p_T$ spectra in $AA$ collisions.
This follows from the
observation~\cite{phobos} $R_{AA}^{N_{part}}\to 1$ for small $p_T$,
i.e. approximate validity of ``participant scaling''. We studied
the issue further by combining the high accuracy $pp$ and $AuAu$ data 
from STAR \cite{star} at 200 GeV with the corresponding nuclear 
data from PHOBOS that extend towards low $p_T\sim
0.03$ GeV. Using the thermal parametrisation (\ref{invfitth}) we find
\begin{equation} 
I_0^{AA}/I_0^{pp}\approx 160\pm 17,
\label{AA/pp}
\end{equation}
which is compatible
with the calculated $N_{part}/2=172\ (\pm15\%)$. Therefore, the normalisation
in the soft limit is consistent with
\begin{equation}
p_T\to 0: \qquad   R_{AA}^{N_{part}}\to 1\quad \text{and}\quad
I_0^{AuAu} \approx  \dfrac{N_{part}}{2}\ I_0^{pp}.
\label{RAApart}
\end{equation}  
We find it quite remarkable that this simple relation works with good
accuracy of about 10\% which corresponds to the accuracy of both 
the measurement and the
theoretical calculation. Note, in particular, that relation
(\ref{RAApart}) is only valid in the limit $p_T\to 0$ and is violated by
about 50\% already at $p_T\sim 0.5$~GeV (see STAR~\cite{star2}); also note that 
the observed particle density for the
``wounded nucleon'' interactions is strongly suppressed 
($\propto \frac{N_{part}}{2}$) 
by the coherence effects and it is about six times smaller than for
the incoherent superposition of nuclear scatterings ($\propto N_{coll}$).  

Within our approach we should relate this normalisation of
particle density in the
soft limit (\ref{RAApart}) to the minimal parton configuration responsible
for the emission of a large wavelength gluon. In this limit with
``participant scaling,'' nucleons of one nucleus
interact with some nucleons of the other nucleus several
times but the outgoing partonic system acts again like a colour octet
source such as if it had scattered only once as in $pp$ collisions. 
One could think of two mechanisms leading to such a result. 

The primary
interaction in a single nucleon-nucleon collision responsible for a
``minimum bias'' event leads to the break up of the scattered 
nucleons dominantly into quark-diquark systems which form colour singlet
or octet states only. One could argue that 
such a semihard collision with $p_T\lapproxeq 1$ GeV may not resolve higher
Fock states in the proton of typically smaller size. Then the soft hadrons 
will not be sensitive to higher
partonic excitations either.\footnote{A
model based on wounded quarks and diquarks has been developed in Ref.
\cite{bialas1}, but 
for the description of $p_T$-integrated rapidity distributions.} 
Another argument can be based on an assumption that at high energies
the multi-gluon exchange related to multiple scatterings of the same
nucleon  dominantly proceeds via the colour octet exchange, while the
higher colour multiplets (such as decuplet or 27-plet) are disfavoured.
Note that in the non-Abelian gauge theory the intercept of the reggeised
 gluon $\alpha_G(0)=1$ (i.e. the
spin of the on-mass-shell gluon is equal to one) while the j-plane
singularities corresponding to the exchange of the higher
colour multiplets are situated to the left of $j=1$~\cite{Bronzan:1977yx}. 
Therefore, the
exchange of higher colour multiplets are asymptotically suppressed. 
It remains a challenge to understand the rapid disappearance with $p_T$ of
the limiting density (\ref{RAApart}).

Another interesting simplified structure appears for particle ratios
in the low $p_T$ region. If we compare \cite{Ochs:2010sv} the $p_T$ dependence of such ratios 
from \epem annihilation ($p_T$ with respect to sphericity axis) with
those from $pp$ scattering \cite{tpc} and also ratios from $pp$ and $AA$
collisions by PHENIX  and STAR we observe the convergence of these spectra
in the region
of low $p_T\lapproxeq 0.5$~GeV while they are quite different otherwise. 
This could be
related to the universal
bremsstrahlung from the colour sources in the considered processes
in this soft limit. 
This observation implies that the soft particles decouple from
equilibration in nuclear collisions. The soft particles in the central
rapidity region are produced first and they stay behind if the system
expands.
 

\begin{footnotesize}

\end{footnotesize}


\end{document}